\documentclass[12pt]{article}
\usepackage{amssym}

\def\case#1#2{{\textstyle{#1\over #2}}}

\title{
Infinite square well and periodic trajectories in classical mechanics}
\author{B. Bagchi $^{a,}$\thanks{E-mail address: bbagchi@cucc.ernet.in}\
, S. Mallik $^a$,  C. Quesne $^{b,}$\thanks{Directeur de recherches FNRS; E-mail address:
cquesne@ulb.ac.be}\\ 
$^a$ {\small Department of Applied Mathematics, University of Calcutta,}\\ 
{\small 92 Acharya Prafulla Chandra Road, Calcutta 700 009, India}\\ 
$^b$ {\small Physique Nucl\'eaire Th\'eorique et Physique
Math\'ematique,  Universit\'e Libre de Bruxelles,} \\ 
{\small Campus de la Plaine CP229, Boulevard~du Triomphe, B-1050 Brussels,
Belgium}}
\date{ }
\begin{document}
\baselineskip=20pt plus 1pt minus 1pt
\maketitle

\begin{abstract} 
We examine the classical problem of an infinite square well by considering Hamilton's
equations in one dimension and Hamilton-Jacobi equation for motion in two dimensions.
We illustrate, by means of suitable examples, the nature of the periodic motion of a particle
trapped inside the well. 
\end{abstract}
%
%
\newpage
\section{Introduction}

The square well potential~\cite{bloch}, also referred to as the square billiard in the
literature~\cite{bulgac}, describes the confinement of a particle trapped in a box with
infinite walls. Being a rather simple system that is solvable in both classical and quantum
mechanics, it has appeared, along with the harmonic oscillator~\cite{goldstein, calkin}, as
an example of natural enquiry in the developments of semiclassical
theories~\cite{mccann, ihra}, supersymmetric quantum mechanics~\cite{sukumar,
devincenzo}, PT symmetry~\cite{znojil01a, znojil01b, bagchi}, and also in models of
coherent states~\cite{antoine}.\par
%
%
At the classical level, the characteristic frequency of the particle in the well is related to
the natural frequency of the harmonic oscillator by a factor $\pi/2$ at the same energy
and amplitude. Indeed choosing the origin at the centre of the well and $x$ axis along its
length, the total energy of a particle of mass $m$ trapped inside the well is given by $E =
(1/2) m v_x^2$, where $v_x$ is the velocity of the particle and the collisions at
the ends are assumed to be perfectly reflective. Note that because of the infinite nature
of the walls the particle cannot go past them. Now, to cover the distance from $x=-a$ to
$x=a$ inside the well, $2a$ being the length of the well, the time $T/2$ taken by the
particle is $T/2 = 2a/v_x$. Writing $T = 2\pi/\omega$, where $\omega$ is the
characteristic frequency that we associate with the particle, it transpires that $v_x = 2a
\omega/\pi$ along with $E = (1/2) m \omega^2 (2a)^2/\pi^2$. Comparing with the
energy formula $E = (1/2) m\Omega^2 a^2$ of the harmonic oscillator corresponding
to an amplitude $a$ and frequency $\Omega$, we arrive at the aforementioned
result.\par
%
%
It may be mentioned that for the quantum mechanical situation, using the
Sommerfeld-Wilson quantization rule~\cite{mavromatis}, namely$\oint p_x\, dx = nh$,
i.e., $\int_{-a}^a mv_x\, dx + \int_a^{-a} (-mv_x)\, dx = nh$, we get $p_x = mv_x =
nh/(4a)$, implying $p_x$ to be quantized. Physically it means that if we decrease $a$
(i.e., resort to squeezing), $p_x$ increases. For the above $p_x$, the quantized energy
levels read $E_n = n^2 h^2/(32 ma^2)$, $n=1$, 2,~\ldots. It follows that the energy
levels of a quantum particle inside the well are entirely discrete, nondegenerate, and not
equispaced as for the case of the harmonic oscillator.\par
%
%
However, contrary to the quantum case, graduate level textbooks are rather brief with
the classical problem of the infinite square well. Our aim in this article is to present a
somewhat detailed mathematical exposition of it in one and two dimensions that brings
out clearly the periodic nature of the trajectories of a particle trapped in it. In section~2,
we deal with the one-dimensional motion starting from Hamilton's equations of motion
and show that the motion of the particle is periodic with period $T = 4ma/p_0$, $p_0$
being its initial momentum. In section~3, we solve the Hamilton-Jacobi equation to
demonstrate how we run into periodic trajectories in two-dimensional systems as well. In
section~4, we illustrate the above results by means of nontrivial but elementary
examples, which to the best of our knowledge, have not appeared before. Finally, in
section~5, a conclusion is presented.\par
%
%
\section{One-dimensional motion}

Let us analyze the classical square well problem in one dimension by starting from
Hamilton's equations. Taking the Hamiltonian $H = p^2/(2m)$ between the two barriers
at $x=-a$ and $x=a$ (note that the particle is free inside the well) and imposing the
initial condition $x_0=0$ at $t=t_0=0$ with the momentum $p_0>0$, we find the
trajectory to be
\begin{equation}
  x = \frac{p_0}{m} t \qquad t > 0.
\end{equation}
No sooner the particle reaches the barrier at $x=a$, which occurs at say $t=t_1$, given
by $t_1= ma/p_0$, it reverses direction and its trajectory then is described by
\begin{equation}
  x = - \frac{p_0}{m} (t-t_1) + a \qquad t > t_1.  \label{eq:1-reverse}
\end{equation}
Obviously, equation~(\ref{eq:1-reverse}) holds until the particle reaches the other barrier
at $x=-a$, which occurs at $t = t_2 = 3ma/p_0$. Then it reverses direction again, its
trajectory being
\begin{equation}
  x = \frac{p_0}{m} (t-t_2) - a \qquad t > t_2.
\end{equation}  
so long as it does not reach $x=a$.\par
%
%
Continuing like this, we may summarize our results for the trajectory as follows
\begin{eqnarray}
  x & = & \frac{p_0}{m} t \qquad t_0 \le t < t_1 \nonumber \\
  x & = & - \frac{p_0}{m} (t - t_{2\nu-1}) + a \qquad t_{2\nu-1} \le t < t_{2\nu}
         \nonumber \\
  x & = & \frac{p_0}{m} (t - t_{2\nu}) - a \qquad t_{2\nu} \le t < t_{2\nu+1}
\end{eqnarray}
where $\nu=1$, 2,~\ldots and $t_n = (2n-1)ma/p_0$, $n=1$, 2,~\ldots.\par
%
%
Hence we conclude that the motion of the particle inside the one-dimensional well is
periodic with period $T = 4ma/p_0$.\par
%
%
\section{Two-dimensional motion}

To enquire into the periodic nature of the trajectories of the particle in two dimensions it
is instructive to start from the Hamilton-Jacobi equation.\par
%
%
Consider the following canonical transformation of the coordinates and momenta
\begin{equation}
  (q, p, t) \to (Q, P, T)
\end{equation}
for two degrees of freedom: $q = (q_1, q_2)$, $p = (p_1, p_2)$, $Q = (Q_1, Q_2)$,
$P = (P_1, P_2)$. The Hamilton-Jacobi equation
\begin{equation}
  H\left(q; \frac{\partial F}{\partial q}; t\right) + \frac{\partial F}{\partial t} = 0 
\end{equation}
is concerned with the canonical transformation from $(q, p)$ to new variables $(\beta,
\alpha)$ that are constant in time:
\begin{equation}
  p = \frac{\partial F(q, \alpha, t)}{\partial q} \qquad \beta = \frac{\partial F(q, \alpha,
  t)}{\partial \alpha}  \label{eq:canonical} 
\end{equation}
where $F$ denotes the type-two generating function, $\beta = (\beta_1, \beta_2)$,
and $\alpha = (\alpha_1, \alpha_2)$.\par
%
%
We consider the time-independent case when $t$ does not appear explicitly in $H$:
\begin{equation}
  H\left(q; \frac{\partial F}{\partial q}\right) + \frac{\partial F}{\partial t} = 0.
  \label{eq:H-J} 
\end{equation}
To solve (\ref{eq:H-J}) the standard procedure is to employ a separation of variables on
$F$:
\begin{equation}
  F = F(q) + T(t).
\end{equation}
This results in
\begin{equation}
  H\left(q, \frac{\partial F}{\partial q}\right) = - \frac{dT}{dt}
\end{equation}
where the left-hand side is a function of $q$ only while the right-hand side is a function
of $t$ only. So each side must be a constant $E$:
\begin{equation}
  H\left(q, \frac{\partial F}{\partial q}\right) = E \qquad \frac{dT}{dt} = -E.
  \label{eq:H-J-E}
\end{equation}
\par
%
%
Consider now a two-dimensional square well with centre at $(0,0)$ and vertices at $(\pm
a, \pm a)$:
\begin{eqnarray}
  V(x, y) & = & 0 \qquad |x| < a {\rm\  and\ }|y| < a \nonumber \\
  & =& \infty \qquad |x| > a {\rm\  or\ }|y| > a. \label{eq:V}
\end{eqnarray}
Inside the well, the first equation of (\ref{eq:H-J-E}) is given by
\begin{equation}
  \frac{1}{2m} \left[\left(\frac{\partial F}{\partial x}\right)^2 + \left(\frac{\partial
  F}{\partial y}\right)^2\right] = E.  \label{eq:H-J-SW} 
\end{equation}
Setting $F = X(x) + Y(y)$ in (\ref{eq:H-J-SW}), it is straightforward to obtain
\begin{equation}
  \left(\frac{dX}{dx}\right)^2 = \alpha_x^2 \qquad \left(\frac{dY}{dy}\right)^2 =
  \alpha_y^2  \label{eq:XY}
\end{equation}
where the constants $\alpha_x^2$ and $\alpha_y^2$ are subjected to the constraint
\begin{equation}
  E = \frac{1}{2m} (\alpha_x^2 + \alpha_y^2).
\end{equation}
Integrating (\ref{eq:XY}) and substituting the result in (\ref{eq:canonical}) yields
\begin{equation}
  p_x = \pm \alpha_x \qquad p_y = \pm \alpha_y.  \label{eq:H-J-sol}
\end{equation}
Equation (\ref{eq:H-J-sol}) is consistent with the fact that as soon as the particle,
moving freely inside the well, approaches the infinite barriers at $x = \pm a$, it reverses
direction.\par
%
%
The action variables can now be evaluated in the following manner
\begin{equation}
  I_x = \oint p_x\, dx = \int_{-a}^a \alpha_x\, dx + \int_a^{-a} (-\alpha_x)\, dx = 4a
  \alpha_x. 
\end{equation}
Similarly we have $I_y = 4a \alpha_y$. As a consequence, $E$ can be expressed as
\begin{equation}
  E = \frac{1}{32ma^2}(I_x^2 + I_y^2).
\end{equation}
The corresponding frequencies are
\begin{equation}
  \omega_x = \frac{\partial E}{\partial I_x} = \frac{I_x}{16ma^2} \qquad
  \omega_y = \frac{\partial E}{\partial I_y} = \frac{I_y}{16ma^2}. 
  \label{eq:frequencies} 
\end{equation}
\par
%
%
We thus see that periodic orbits occur when the periods of motion $T_x$ and $T_y$
along the $x$ and $y$ axes are such that
\begin{equation}
  \frac{T_y}{T_x} = \frac{\omega_x}{\omega_y} = \frac{I_x}{I_y} = \frac{n_x}{n_y}
  \label{eq:T-ratio}\end{equation}
where $n_x$ and $n_y$ are integers. In other words, the particle always returns to its
starting point with its initial velocity with a period
\begin{equation}
  T = n_x T_x = n_y T_y.  \label{eq:T}
\end{equation}
Equation (\ref{eq:T-ratio}) is the central result for the two-dimensional case,
which we illustrate below by means of some simple examples.\par
%
%
\section{Illustrations}

Let a particle, under the influence of the potential (\ref{eq:V}), start from the centre
with a momentum making an angle $\theta$ with the $x$ axis, i.e.,
\begin{equation}
  t_0=0 \qquad x_0 = y_0 = 0 \qquad \frac{p_{y_0}}{p_{x_0}} = \tan\theta.
  \label{eq:initial}
\end{equation}
If $\tan\theta$ is a rational number, i.e., $\tan\theta = n_y/n_x$ where $n_x$ and
$n_y$ are integers, then the motion is periodic with a period $T$ given by (\ref{eq:T}). To
show how equations (\ref{eq:T}) and (\ref{eq:initial}) may work in practice, we consider
two concrete examples.\par
%
%
\subsection{\boldmath Example 1: $\tan\theta = 1/2$ or $n_x = 2$, $n_y = 1$}

Here we notice that during the first period, the particle takes the following path (see
figure~1):
\begin{itemize}
\item $x = (p_{x_0}/m) t$, $y = (p_{y_0}/m) t$ if $t_0 = 0 \le t < t_1 = ma/p_{x_0}$.
At
$t = t_1$, $x_1 = a$, $y_1 = a/2$.

\item $x = - (p_{x_0}/m) (t - t_1) + a$, $y = (p_{y_0}/m) (t - t_1) + a/2$ if $t_1 \le t
< t_2 = 2t_1$. At $t = t_2$, $x_2 = 0$, $y_2 = a$. 

\item $x = - (p_{x_0}/m) (t - t_2)$, $y = - (p_{y_0}/m) (t - t_2) + a$ if $t_2 \le t
< t_3 = 3t_1$. At $t = t_3$, $x_3 = - a$, $y_3 = a/2$. 

\item $x = (p_{x_0}/m) (t - t_3) - a$, $y = - (p_{y_0}/m) (t - t_3) + a/2$ if $t_3 \le t
< t_5 = 5t_1$. At $t = t_5$, $x_5 = a$, $y_5 = - a/2$. Note that at $t = t_4 =
4t_1$, the particle goes through the origin ($x_4 = y_4 = 0$), but the components of
its momentum are different from those at $t = t_0$, namely $(p_{x_0}, - p_{y_0})$
instead of $(p_{x_0}, p_{y_0})$.

\item $x = - (p_{x_0}/m) (t - t_5) + a$, $y = - (p_{y_0}/m) (t - t_5) - a/2$ if $t_5 \le
t < t_6 = 6t_1$. At $t = t_6$, $x_6 = 0$, $y_6 = -a$.  

\item $x = - (p_{x_0}/m) (t - t_6)$, $y = (p_{y_0}/m) (t - t_6) - a$ if $t_6 \le t
< t_7 = 7t_1$. At $t = t_7$, $x_7 = - a$, $y_7 = -a/2$.  

\item $x = (p_{x_0}/m) (t - t_7) - a$, $y = (p_{y_0}/m) (t - t_7) - a/2$ if $t_7 \le t
< t_8 = 8t_1$. At $t = t_8$, $x_8 = y_8 = 0$ and the particle has the same
momentum as at $t = t_0$.
\end{itemize}
\par
%
%
The period is therefore $T = t_8 = 8t_1 = 8ma/p_{x_0}$. One also notes that the periods
$T_x$ and $T_y$ are 
\begin{equation}
  T_x = t_4 = 4t_1 = \frac{4ma}{p_{x_0}} \qquad T_y = t_8 = 8t_1 =
  \frac{8ma}{p_{x_0}} 
\end{equation}
so that equation (\ref{eq:T}) becomes $T = 2T_x = T_y$, as it should be.\par
%
%
The general solution of the equations of motion may be written as
\begin{eqnarray}
  x & = & \frac{p_{x_0}}{m} t \qquad y = \frac{p_{y_0}}{m} t \qquad {\rm if\ } t_0 \le t
        < t_1 \nonumber \\
  x & = & - \frac{p_{x_0}}{m} (t - t_{8\nu-7}) + a \qquad y = \frac{p_{y_0}}{m} (t -
        t_{8\nu-7}) + \frac{a}{2}\qquad {\rm if\ } t_{8\nu-7} \le t < t_{8\nu-6}
        \nonumber \\
  x & = & - \frac{p_{x_0}}{m} (t - t_{8\nu-6}) \qquad y = - \frac{p_{y_0}}{m} (t -
        t_{8\nu-6}) + a \qquad {\rm if\ } t_{8\nu-6} \le t < t_{8\nu-5}
        \nonumber \\
  x & = & \frac{p_{x_0}}{m} (t - t_{8\nu-5}) - a \qquad y = - \frac{p_{y_0}}{m} (t -
        t_{8\nu-5}) + \frac{a}{2}\qquad {\rm if\ } t_{8\nu-5} \le t < t_{8\nu-3}
        \nonumber \\
  x & = & - \frac{p_{x_0}}{m} (t - t_{8\nu-3}) + a \qquad y = - \frac{p_{y_0}}{m} (t -
        t_{8\nu-3}) - \frac{a}{2}\qquad {\rm if\ } t_{8\nu-3} \le t < t_{8\nu-2}
        \nonumber \\
  x & = & - \frac{p_{x_0}}{m} (t - t_{8\nu-2}) \qquad y = \frac{p_{y_0}}{m} (t -
        t_{8\nu-2}) - a \qquad {\rm if\ } t_{8\nu-2} \le t < t_{8\nu-1}
        \nonumber \\
  x & = & \frac{p_{x_0}}{m} (t - t_{8\nu-1}) - a \qquad y = \frac{p_{y_0}}{m} (t -
        t_{8\nu-1}) - \frac{a}{2}\qquad {\rm if\ } t_{8\nu-1} \le t < t_{8\nu+1}
\end{eqnarray}
where $\nu = 1$, 2,~\ldots, $t_0 = 0$, $t_1 = ma/p_{x_0}$, $t_n = n t_1$, $n=1$,
2,~\ldots.\par
%
%
\subsection{\boldmath Example 2: $\tan\theta = 2/3$ or $n_x = 3$, $n_y = 2$}

{}For this case, the particle goes along the following path (see figure~2):
\begin{itemize}
\item $x = (p_{x_0}/m) t$, $y = (p_{y_0}/m) t$ if $t_0 = 0 \le t < t_1 = ma/p_{x_0}$.
At $t = t_1$, $x_1 = a$, $y_1 = 2a/3$.

\item $x = - (p_{x_0}/m) (t - t_1) + a$, $y = (p_{y_0}/m) (t - t_1) + 2a/3$ if $t_1 \le
t < t_2 = 3t_1/2$. At $t = t_2$, $x_2 = a/2$, $y_2 = a$. 

\item $x = - (p_{x_0}/m) (t - t_2)$ + a/2, $y = - (p_{y_0}/m) (t - t_2) + a$ if $t_2 \le t
< t_3 = 3t_1$. At $t = t_3$, $x_3 = - a$, $y_3 = 0$. 

\item $x = (p_{x_0}/m) (t - t_3) - a$, $y = - (p_{y_0}/m) (t - t_3)$ if $t_3 \le t
< t_4 = 9t_1/2$. At $t = t_4$, $x_4 = a/2$, $y_4 = - a$. 

\item $x = (p_{x_0}/m) (t - t_4) + a/2$, $y = (p_{y_0}/m) (t - t_4) - a$ if $t_4 \le t
< t_5 = 5t_1$. At $t = t_5$, $x_5 = a$, $y_5 = - 2a/3$. 

\item $x = - (p_{x_0}/m) (t - t_5) + a$, $y = (p_{y_0}/m) (t - t_5) - 2a/3$ if $t_5 \le
t < t_7 = 7t_1$. At $t = t_7$, $x_7 = - a$, $y_7 = 2a/3$. Note that at $t = t_6 =
6t_1$, the particle goes through the origin ($x_6 = y_6 = 0$), but the components of
the momentum are different from those at $t = t_0$, namely $(- p_{x_0}, p_{y_0})$
instead of $(p_{x_0}, p_{y_0})$.

\item $x = (p_{x_0}/m) (t - t_7) - a$, $y = (p_{y_0}/m) (t - t_7)  + 2a/3$ if $t_7 \le t
< t_8 = 15t_1/2$. At $t = t_8$, $x_8 = - a/2$, $y_8 = a$.  

\item $x = (p_{x_0}/m) (t - t_8) - a/2$, $y = - (p_{y_0}/m) (t - t_8) + a$ if $t_8 \le
t < t_9 = 9t_1$. At $t = t_9$, $x_9 = a$, $y_9 = 0$.

\item $x = - (p_{x_0}/m) (t - t_9) + a$, $y = - (p_{y_0}/m) (t - t_9)$ if $t_9 \le
t < t_{10} = 21t_1/2$. At $t = t_{10}$, $x_{10} = - a/2$, $y_{10} = - a$.

\item $x = - (p_{x_0}/m) (t - t_{10}) - a/2$, $y = (p_{y_0}/m) (t - t_{10}) - a$ if
$t_{10} \le t < t_{11} = 11t_1$. At $t = t_{11}$, $x_{11} = - a$, $y_{11} = - 2a/3$.

\item $x = (p_{x_0}/m) (t - t_{11}) - a$, $y = (p_{y_0}/m) (t - t_{11}) - 2a/3$ if
$t_{11} \le t < t_{12} = 12t_1$. At $t = t_{12}$, $x_{12} = y_{12} = 0$ and the
particle has the same momentum as at $t = t_0$.
\end{itemize}
\par
%
%
The period is therefore $T = t_{12} = 12t_1 = 12ma/p_{x_0}$ along with
\begin{equation}
  T_x = t_5 - t_1 = 4t_1 = \frac{4ma}{p_{x_0}} \qquad T_y = t_8 - t_2 = 6t_1 =
  \frac{6ma}{p_{x_0}}. 
\end{equation}
Equation (\ref{eq:T}) thus becomes $T = 3T_x = 2T_y$, as it should be.
\par
%
%
The general solution of the equations of motion may be written as
\begin{eqnarray}
  x & = & \frac{p_{x_0}}{m} t \qquad y = \frac{p_{y_0}}{m} t \qquad {\rm if\ } t_0 \le t
        < t_1 \nonumber \\
  x & = & - \frac{p_{x_0}}{m} (t - t_{12\nu-11}) + a \qquad y = \frac{p_{y_0}}{m} (t -
        t_{12\nu-11}) + \frac{2a}{3}\qquad {\rm if\ } t_{12\nu-11} \le t < t_{12\nu-10}
        \nonumber \\
  x & = & - \frac{p_{x_0}}{m} (t - t_{12\nu-10}) + \frac{a}{2} \qquad y =
        - \frac{p_{y_0}}{m} (t - t_{12\nu-10}) + a\qquad {\rm if\ } t_{12\nu-10}
        \le t < t_{12\nu-9} \nonumber \\ 
  x & = & \frac{p_{x_0}}{m} (t - t_{12\nu-9}) - a \qquad y = - \frac{p_{y_0}}{m} (t -
        t_{12\nu-9}) \qquad {\rm if\ } t_{12\nu-9} \le t < t_{12\nu-8}
        \nonumber \\
  x & = & \frac{p_{x_0}}{m} (t - t_{12\nu-8}) + \frac{a}{2} \qquad y =
        \frac{p_{y_0}}{m} (t - t_{12\nu-8}) - a\qquad {\rm if\ } t_{12\nu-8} \le t <
        t_{12\nu-7} \nonumber \\
  x & = & - \frac{p_{x_0}}{m} (t - t_{12\nu-7}) + a \qquad y = \frac{p_{y_0}}{m} (t -
        t_{12\nu-7}) - \frac{2a}{3}\qquad {\rm if\ } t_{12\nu-7} \le t < t_{12\nu-5}
        \nonumber \\
  x & = & \frac{p_{x_0}}{m} (t - t_{12\nu-5}) - a \qquad y = \frac{p_{y_0}}{m} (t -
        t_{12\nu-5}) + \frac{2a}{3} \qquad {\rm if\ } t_{12\nu-5} \le t < t_{12\nu-4}
        \nonumber \\
  x & = & \frac{p_{x_0}}{m} (t - t_{12\nu-4}) - \frac{a}{2} \qquad y =
         - \frac{p_{y_0}}{m} (t - t_{12\nu-4}) + a \qquad {\rm if\ } t_{12\nu-4} \le t <
        t_{12\nu-3} \nonumber \\
  x & = & - \frac{p_{x_0}}{m} (t - t_{12\nu-3}) + a \qquad y = - \frac{p_{y_0}}{m} (t -
        t_{12\nu-3}) \qquad {\rm if\ } t_{12\nu-3} \le t < t_{12\nu-2} \nonumber \\
  x & = & - \frac{p_{x_0}}{m} (t - t_{12\nu-2}) - \frac{a}{2} \qquad y =
         \frac{p_{y_0}}{m} (t - t_{12\nu-2}) - a \qquad {\rm if\ } t_{12\nu-2} \le t <
        t_{12\nu-1} \nonumber \\
  x & = & \frac{p_{x_0}}{m} (t - t_{12\nu-1}) - a \qquad y = \frac{p_{y_0}}{m} (t -
        t_{12\nu-1}) - \frac{2a}{3} \qquad {\rm if\ } t_{12\nu-1} \le t < t_{12\nu+1} 
\end{eqnarray}
where $\nu = 1$, 2,~\ldots, $t_0 = 0$, $t_1 = ma/p_{x_0}$, and 
\begin{eqnarray*}
  t_n & = & n t_1 \qquad {\rm if\ } n \ne 12\nu-10, 12\nu-8, 12\nu-4, 12\nu-2 \\
  & = & (n - \case{1}{2}) t_1 \qquad {\rm if\ } n = 12\nu-10, 12\nu-4 \\
  & = & (n + \case{1}{2}) t_1 \qquad {\rm if\ } n = 12\nu-8, 12\nu-2. 
\end{eqnarray*}
\par
%
%
{}Finally, $T_x$ and $T_y$ are given by
\begin{equation}
  T_x = \frac{4ma}{p_{x_0}} \qquad T_y = \frac{4ma}{p_{y_0}}.
\end{equation}
The corresponding frequencies are
\begin{eqnarray}
  \omega_x & = & \frac{p_{x_0}}{4ma} = \frac{\alpha_x}{4ma} = \frac{I_x}{16ma^2} 
        \nonumber \\
  \omega_y & = & \frac{p_{y_0}}{4ma} = \frac{\alpha_y}{4ma} = \frac{I_y}{16ma^2} 
\end{eqnarray}
in agreement with the result (\ref{eq:frequencies}).\par
%
%
\section{Conclusion}

The motion of a particle trapped inside a square well with infinite barriers is well known to
be periodic. While most textbooks leave the issue at that, we have, in this article, pursued
the problem a little further by taking a closer look at the basic equations and enquiring as
to how these may work in practice. To this end, we have examined Hamilton's equations
in one dimension and Hamilton-Jacobi equation for motion in two dimensions. We have
then constructed a couple of insightful examples to demonstrate how the periodicity of
trajectories takes place.\par
%
%
 
%
%
\newpage
\section*{Figure captions}

{\bf Figure 1.} Motion of a particle inside a two-dimensional square well in the case where
$\tan \theta = 1/2$.

\noindent
{\bf Figure 2.} Motion of a particle inside a two-dimensional square well in the case where
$\tan \theta = 2/3$.
%
%
\newpage
\begin{picture}(160,120)(-15,0)

\put(60,-15){\vector(0,1){150}}
\put(-15,60){\vector(1,0){150}}
\put(135,60){\makebox(5,0)[r]{\Large x}}
\put(60,135){\makebox(0,5)[t]{\Large y}}

\put(0,0){\thicklines\line(0,1){120}}
\put(120,0){\thicklines\line(0,1){120}}
\put(0,0){\thicklines\line(1,0){120}}
\put(0,120){\thicklines\line(1,0){120}}
\put(-4,60){\makebox(0,4)[t]{\Large -a}}
\put(123,60){\makebox(0,4)[t]{\Large a}}

\put(60,60){\thicklines\vector(2,1){15}}
\put(75,67.5){\thicklines\line(2,1){45}}
\put(120,90){\thicklines\vector(-2,1){15}}
\put(105,97.5){\thicklines\line(-2,1){45}}
\put(60,120){\thicklines\vector(-2,-1){15}}
\put(45,112.5){\thicklines\line(-2,-1){45}}
\put(0,90){\thicklines\vector(2,-1){15}}
\put(15,82.5){\thicklines\vector(2,-1){60}}
\put(75,52.5){\thicklines\line(2,-1){45}}
\put(120,30){\thicklines\vector(-2,-1){15}}
\put(105,22.5){\thicklines\line(-2,-1){45}}
\put(60,0){\thicklines\vector(-2,1){15}}
\put(45,7.5){\thicklines\line(-2,1){45}}
\put(0,30){\thicklines\vector(2,1){15}}
\put(15,37.5){\thicklines\line(2,1){45}}


\put(62,62){\makebox(3,0)[b]{\Large $t_0$}}
\put(123,90){\makebox(3,0)[r]{\Large $t_1$}}
\put(62,121){\makebox(3,0)[b]{\Large $t_2$}}
\put(-4,90){\makebox(3,0)[r]{\Large $t_3$}}
\put(56,62){\makebox(3,0)[b]{\Large $t_4$}}
\put(123,30){\makebox(3,0)[r]{\Large $t_5$}}
\put(62,-1){\makebox(3,0)[t]{\Large $t_6$}}
\put(-4,30){\makebox(3,0)[r]{\Large $t_7$}}
\put(56,53){\makebox(3,0)[b]{\Large $t_8$}}

\end{picture}

\vspace{5cm}
\centerline{Figure 1}
%
%
\newpage
\begin{picture}(160,120)(-15,0)

\put(60,-15){\vector(0,1){150}}
\put(-15,60){\vector(1,0){150}}
\put(135,60){\makebox(5,0)[r]{\Large x}}
\put(60,135){\makebox(0,5)[t]{\Large y}}

\put(0,0){\thicklines\line(0,1){120}}
\put(120,0){\thicklines\line(0,1){120}}
\put(0,0){\thicklines\line(1,0){120}}
\put(0,120){\thicklines\line(1,0){120}}
\put(62,-2){\makebox(3,0)[t]{\Large -a}}
\put(62,122){\makebox(3,0)[b]{\Large a}}

\put(60,60){\thicklines\vector(3,2){15}}
\put(75,70){\thicklines\line(3,2){45}}
\put(120,100){\thicklines\vector(-3,2){7.5}}
\put(112.5,105){\thicklines\line(-3,2){22.5}}
\put(90,120){\thicklines\vector(-3,-2){22.5}}
\put(67.5,105){\thicklines\line(-3,-2){67.5}}
\put(0,60){\thicklines\vector(3,-2){22.5}}
\put(22.5,45){\thicklines\line(3,-2){67.5}}
\put(90,0){\thicklines\vector(3,2){7.5}}
\put(97.5,5){\thicklines\line(3,2){22.5}}
\put(120,20){\thicklines\vector(-3,2){15}}
\put(105,30){\thicklines\line(-3,2){45}}
\put(60,60){\thicklines\vector(-3,2){15}}
\put(45,70){\thicklines\line(-3,2){45}}
\put(0,100){\thicklines\vector(3,2){7.5}}
\put(7.5,105){\thicklines\line(3,2){22.5}}
\put(30,120){\thicklines\vector(3,-2){22.5}}
\put(52.5,105){\thicklines\line(3,-2){67.5}}
\put(120,60){\thicklines\vector(-3,-2){22.5}}
\put(97.5,45){\thicklines\line(-3,-2){67.5}}
\put(30,0){\thicklines\vector(-3,2){7.5}}
\put(22.5,5){\thicklines\line(-3,2){22.5}}
\put(0,20){\thicklines\vector(3,2){15}}
\put(15,30){\thicklines\line(3,2){45}}

\put(62,64){\makebox(3,0)[b]{\Large $t_0$}}
\put(123,100){\makebox(3,0)[r]{\Large $t_1$}}
\put(90,121){\makebox(3,0)[br]{\Large $t_2$}}
\put(-5,61){\makebox(3,0)[b]{\Large $t_3$}}
\put(90,-1){\makebox(3,0)[tr]{\Large $t_4$}}
\put(123,20){\makebox(3,0)[r]{\Large $t_5$}}
\put(62,56){\makebox(3,0)[t]{\Large $t_6$}}
\put(-5,100){\makebox(3,0)[]{\Large $t_7$}}
\put(30,121){\makebox(3,0)[br]{\Large $t_8$}}
\put(123,61){\makebox(3,0)[b]{\Large $t_9$}}
\put(30,-1){\makebox(3,0)[tr]{\Large $t_{10}$}}
\put(-5,20){\makebox(3,0)[]{\Large $t_{11}$}}
\put(55,56){\makebox(3,0)[t]{\Large $t_{12}$}}

\end{picture}

\vspace{5cm}
\centerline{Figure 2}


\newpage
\begin{thebibliography}{99}

\bibitem{bloch} Bloch S C 1997 {\em Introduction to Classical and Quantum Harmonic
Oscillators} (New York: Wiley)

\bibitem{bulgac} Bulgac A and Magierski P 1999 Eigenstates for billiards of arbitrary
shapes {\em Preprint} physics/9902057

\bibitem{goldstein} Goldstein H 1980 {\em Classical Mechanics} (Reading, MA:
Addison-Wesley)

\bibitem{calkin} Calkin M G 1996 {\em Lagrangian and Hamiltonian Mechanics}
(Singapore: World Scientific)

\bibitem{mccann} McCann E and Richter K 1998 {\em Europhys.\ Lett.} {\bf 43} 241

\bibitem{ihra} Ihra W, Leadbeater M, Vega J L and Richter K 2001 {\em Eur.\ Phys.\ J.}
B {\bf 21} 425

\bibitem{sukumar} Sukumar C V 1985 {\em J.\ Phys.\ A: Math.\ Gen.} {\bf 18} L57

\bibitem{devincenzo} De Vincenzo S and Alonso V 2002 {\em Phys.\ Lett.} {\bf 298A}
98

\bibitem{znojil01a} Znojil M 2001 {\em Phys.\ Lett.} {\bf 285A} 7

\bibitem{znojil01b} Znojil M and Levai G 2001 {\em Mod.\ Phys.\ Lett.} A {\bf 16} 2273

\bibitem{bagchi} Bagchi B, Mallik S and Quesne C 2002 PT-symmetric square well and
the associated SUSY hierarchies {\em Preprint} quant-ph/0205003

\bibitem{antoine} Antoine J-P, Gazeau J-P, Monceau P, Klauder J R and Penson K A 2001
{\em J.\ Math.\ Phys.} {\bf 42} 2349

\bibitem{mavromatis} Mavromatis H A 1987 {\em Exercises in Quantum Mechanics}
(Dordrecht: Reidel) 

\end {thebibliography}
\end{document}